\documentclass[12pt]{iopart}
\usepackage{graphicx}
\usepackage{bm}
\usepackage{bbold}
\expandafter\let\csname equation*\endcsname\relax
\expandafter\let\csname endequation*\endcsname\relax
\usepackage{amsmath}
\usepackage{amsfonts}
\usepackage{amssymb}
\usepackage{amscd}
\usepackage{amstext}

\usepackage{url, hyperref}

\begin{document}

\title{Many-body decoherence dynamics and optimised operation of a single-photon switch}

\author{C. R. Murray$^1$, A. V. Gorshkov$^2$ and T. Pohl$^1$}
\address{$^1$ Max Planck Institute for the Physics of Complex Systems, N\"othnitzer Stra\ss e 38, 01187 Dresden, Germany}
\address{$^2$ Joint Quantum Institute and Joint Center for Quantum Information and Computer Science, NIST/University of Maryland, College Park, Maryland 20742, USA}

\begin{abstract}
We develop a theoretical framework to characterize the decoherence dynamics due to multi-photon scattering in an all-optical switch based on Rydberg atom induced nonlinearities. By incorporating the knowledge of this decoherence process into optimal photon storage and retrieval strategies, we establish optimised switching protocols for experimentally relevant conditions, and evaluate the corresponding limits in the achievable fidelities. Based on these results we work out a simplified description that reproduces recent experiments [arXiv:1511.09445] and provides a new interpretation in terms of many-body decoherence involving multiple incident photons and multiple gate excitations forming the switch. Aside from offering insights into the operational capacity of realistic photon switching capabilities, our work provides a complete description of spin wave decoherence in a Rydberg quantum optics setting, and has immediate relevance to a number of further applications employing photon storage in Rydberg media.
\end{abstract}

\section{Introduction}

An all-optical switch is a device through which the transmission of one optical `target' field can be regulated by the application a second optical `gate' field \cite{Miller2010}. Recently, significant efforts have been directed to reaching the fundamental limit of such a device, in which only a single incoming gate photon is sufficient to switch the target field transmission \cite{Chang2007, OShea2013, Shomroni2014, Tiecke2014, Hwang2009, Bose2012, Volz2012, Bajcsy2009, Chen2013, Reiserer2013}. Such a capability is enticing as it would enable a range of novel functionalities, such as photon multiplexing \cite{Migdall2002, Collins2013}, photonic quantum logic \cite{Nielsen2000, Briegel1998} or nondestructive photo-detection \cite{Gunter2012, Reiserer2013, Gorniaczyk2014, Tiarks2014}.

\begin{figure}[!b]
\hspace*{2.48cm}\includegraphics[width=0.84\columnwidth]{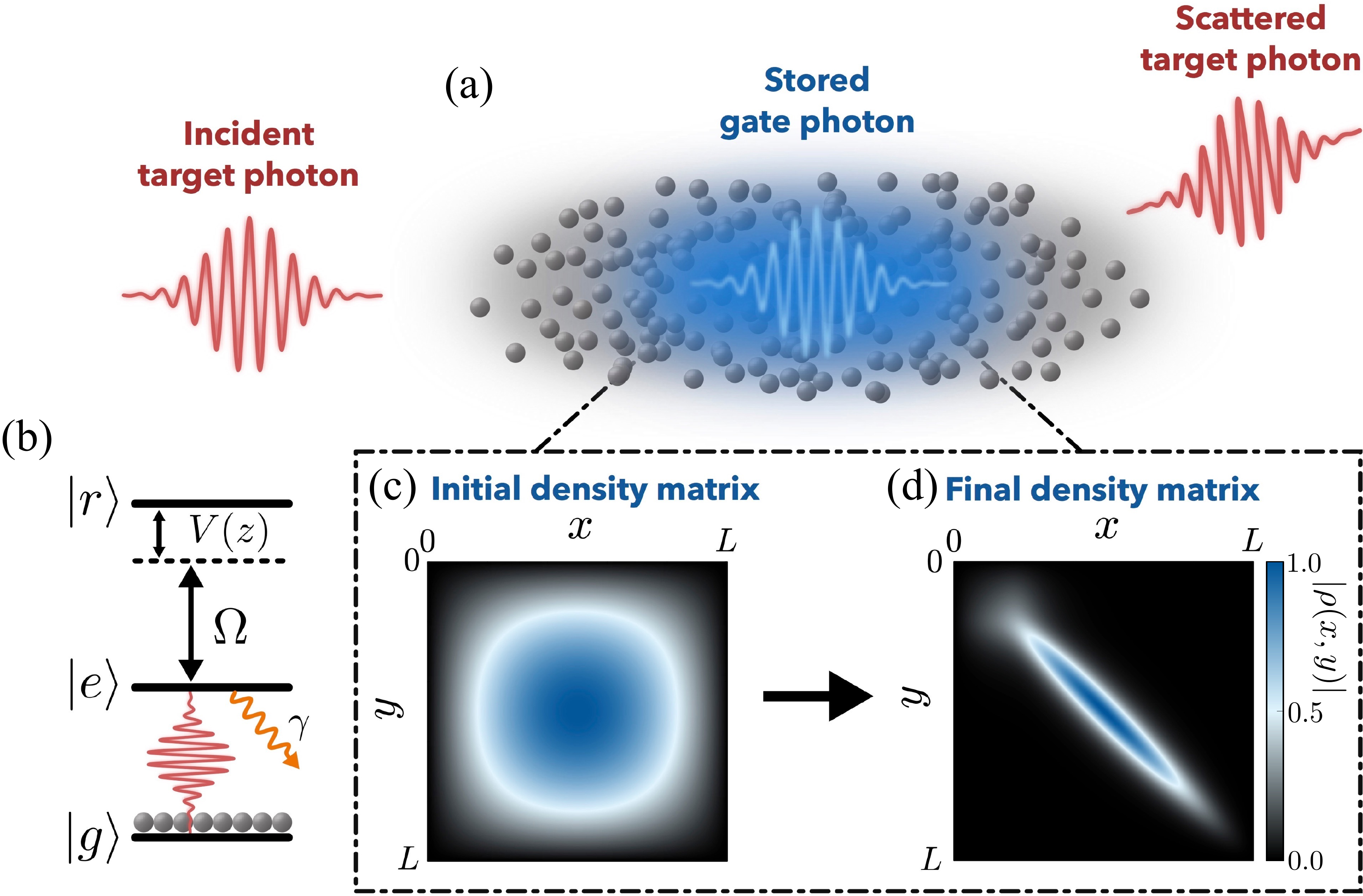}
\caption{\label{fig: schematic} (a) By storing a single `gate' photon (blue) in an atomic medium as a Rydberg spin wave excitation, the transmission of all subsequently incident `target' photons (red) under Rydberg EIT conditions (b) is strongly suppressed. Target photon scattering in this case amounts to projective measurements of the stored spin wave state, causing it to decohere into a statistical mixture of localized excitations. The spatial density matrix, $|\rho(x,y)|$, of the stored spin wave before and after scattering a single photon is shown in (c) and (d) respectively.}
\end{figure}

One way to achieve the large optical nonlinearities \cite{Murray2016,Firstenberg2016} required for single photon switching is by means of electromagnetically induced transparency (EIT) \cite{Fleischhauer2005} with strongly interacting Rydberg states \cite{Saffman2010} in atomic ensembles (see Refs.~\cite{Friedler2005, Pritchard2010, Gorshkov2011, Ates2011, Dudin2012, Peyronel2012, Firstenberg2013, Gorshkov2013, Maxwell2013, Bienias2014,Tresp2015}). The dissipative optical nonlinearities available with this approach \cite{Pritchard2010, Sevincli2011b, Ates2011, Sevincli2011, Dudin2012, Peyronel2012, Gorshkov2013} provide a novel mechanism for single-photon detection \cite{Baur2014}, generation \cite{Gorshkov2013} and substraction \cite{Tresp2016} as well as classical switching capabilities, as recently demonstrated in Refs.~\cite{Baur2014, Gorniaczyk2014, Tiarks2014, Gorniaczyk2015}. Here, the storage of a single gate photon in the medium \cite{Fleischhauer2002, Gorshkov2007, Novikova2007} as a collective Rydberg spin wave excitation is used to cause scattering of all subsequently applied target photons that would otherwise be transmitted (see Fig.~\ref{fig: schematic}). However, the photon scattering in this case amounts to projective measurements of the stored spin wave state, resulting in its decoherence, as shown in Figs.~\ref{fig: schematic}(c) and (d). This has a detrimental effect on the ability to finally retrieve the gate photon, a crucial capability for most practical applications involving switching. Decoherence due to a single target photon has been considered in the asymptotic limit of high atomic densities neglecting photon transmission \cite{Li2015a}, and reduced retrieval efficiencies with increasing target-field intensities have recently been observed experimentally \cite{Gorniaczyk2015}. Yet, a complete picture of the scattering-induced decoherence and its effect on practical multi-photon switching capabilities has not emerged thus far.

Here, we provide such an understanding by deriving an exact solution to the many-body decoherence dynamics of stored gate photons due to interactions with multiple target photons. Incorporating the knowledge of the revealed decoherence physics into optimal photon storage and retrieval strategies \cite{Gorshkov2007, Gorshkov2007a}, we determine and assess the maximum overall switch performance. While photon storage in a short medium \cite{Li2015a} is expected to offer best protection against decoherence, we show that this is not the universally optimal approach to photon switching, particularly for parameter regimes accessible in current experiments \cite{Baur2014, Gorniaczyk2014, Tiarks2014, Gorniaczyk2015}. 

\section{Switch operation}
Outlining the switching protocol in more detail, it is assumed that a single gate photon is first stored \cite{Fleischhauer2000, Gorshkov2007, Novikova2007} as a collectively excited Rydberg state $|r^{\prime}\rangle$ of an atomic ensemble of length $L$, as illustrated in Fig.~\ref{fig: schematic}(a). Subsequently to this, the target field is made to propagate through the medium under EIT conditions involving another long-lived Rydberg state $|r\rangle$ and a low-lying intermediate state $|e\rangle$ that decays with a rate constant $2 \gamma$ [see Fig.~\ref{fig: schematic}(b)]. Low-loss propagation is ensured if the frequency components of the target pulse fit within the EIT spectral window $\sim \Gamma_{\text{EIT}}/ \sqrt{d}$, where $\Gamma_{\text{EIT}}=\Omega^2/\gamma$ is the single-atom EIT linewidth,  $\Omega$ is the Rabi frequency of the classical control field that couples $|e\rangle$ and $|r\rangle$ and $2 d = 2 g^2 n L/(\gamma c)$ is the optical depth of the medium. Here, $g$ is the light-matter coupling strength of the target photons, $n$ is the atomic density, and $c$ is the speed of light.  The van der Waals interaction between $|r\rangle$ and $|r^{\prime}\rangle$ at positions $z$ and $z^{\prime}$, however, results in a spatially dependent level shift $V(z-z^{\prime})=C_6/|z - z^{\prime}|^6$ for the state $|r\rangle$ that ultimately breaks EIT conditions for target photons within a blockade radius $z_b=(C_6/\Gamma_{\text{EIT}})^{1/6}$ \cite{Gorshkov2011} of the stored excitation. This blockade effect essentially exposes a locally absorbing two-level medium composed of $|g\rangle$ and $|e\rangle$ over a spatial extent $2z_b$. The target field then experiences an exponential amplitude attenuation of approximately $\exp [ - 2 d_b ]$ \cite{Gorshkov2011} as it propagates through this region, where $2 d_b=2g^2 n z_b/(\gamma c)$ is the optical depth per blockade radius. In this case, it is clear that large values of $d_b$ are required to significantly suppress the target field transmission in order to achieve efficient switching.

\section{Spin wave decoherence dynamics}
In formally describing the system evolution, we introduce the slowly varying bosonic operator $\hat{\mathcal{E}}^{\dagger}(z)$ that creates a photon in the target field at position $z$, and similarly introduce the operators $\hat{P}^{\dagger}(z)$, $\hat{S}^{\dagger}(z)$ and $\hat{C}^{\dagger}(z)$ to describe the creation of collective atomic excitations in $|e\rangle$, $|r\rangle$ and $|r^{\prime}\rangle$, respectively. The field operators obey Bosonic commutation relations, $[\hat{\mathcal{E}}(z),\hat{\mathcal{E}}^\dagger(z)] = \delta(z-z^\prime)$, etc. In a one-dimensional continuum approximation with homogeneous atomic density, the EIT propagation dynamics of the target field can then be characterized in a rotating frame according to the following set of Heisenberg equations of motion \cite{Gorshkov2011}
\begin{align}
\label{eq: EIT equation 1}
\partial_t \hat{\mathcal{E}}(z, t) & = -c \partial_z \hat{\mathcal{E}}(z, t) - i G \hat{P}(z, t), \\
\label{eq: EIT equation 2}
\partial_t \hat{P}(z, t) & = -i G \hat{\mathcal{E}}(z, t) - i \Omega \hat{S}(z ,t) - \gamma \hat{P}(z, t), \\
\label{eq: EIT equation 3}
\partial_t \hat{S}(z, t) & = - i \Omega \hat{P}(z, t) - i \int_0^L dz^{\prime} V(z-z^{\prime}) \hat{\rho}(z^{\prime}, z^{\prime}) \hat{S}(z, t). 
\end{align}
where we have introduced the collectively enhanced atom-photon coupling $G=g\sqrt{n}$. Since we will only be evaluating normal-ordered expectation values, the vacuum Langevin noise associated with $\gamma$ will not contribute and is thus omitted.  We have furthermore introduced the operator $\hat{\rho}(x, y) = \hat{C}^{\dagger}(x) \hat{C}(y)$, which will later be used to define the elements of the stored spin wave density matrix. In the last equation, the van der Waals interaction between the Rydberg spin wave, described by $\hat{S}(z)$, and the stored Rydberg density, described by $\hat{\rho}(z^{\prime}, z^{\prime})$, is what mediates the effective interaction between the target photon field and the stored gate photon. We further assume ideal switching conditions in which we neglect the self-interactions between target photons that may arise from mutual van der Waals interactions between associated Rydberg atoms in state $|r\rangle$. This approximation is justified provided the intensity of the input field is sufficiently weak \cite{Pritchard2010,Sevincli2011}, and further benefits from choosing the Rydberg states such that the $|r\rangle - |r^{\prime}\rangle$ interactions are enhanced relative to the $|r\rangle - |r\rangle$ interactions \cite{Saffman2009}, e.g., by working close to a interstate F\"orster resonance \cite{Gorniaczyk2014,Tiarks2014}. Finally, the governing equation of motion for $\hat{\rho}(x, y)$ can be written as
\begin{equation}
\label{eq: Density matrix equation of motion}
i \partial_t  \hat{\rho}(x, y, t) = \int dz \left[V(z-y) - V(z-x) \right]  \hat{S}^{\dagger}(z) \hat{\rho}(x, y, t) \hat{S}(z).
\end{equation}
Firstly, one finds that the diagonal elements of $\hat{\rho}(x, y, t)$, i.e. the local spin wave population, are time independent, reflecting the fact that $|r^{\prime}\rangle$ is not laser coupled while the target photons propagate. However, its off-diagonal elements, i.e. the spin wave coherence, are strongly influenced by target photon scattering, as we shall now investigate.

To solve the scattering induced decoherence, let us proceed by considering the state $|\Psi_n\rangle$ in the Heisenberg picture, containing $n$ photons in the mode $\hat{\mathcal{E}}^{\dagger}(z)$ within a temporal envelope $h(t)$ ($\int{\rm d}t|h(t)|^2=1$) and one stored spin wave excitation in the mode $\hat{C}^{\dagger}(z)$ with spatial profile $\mathcal{C}(z)$ ($\int{\rm d}z|\mathcal{C}(z)|^2=1$). Formally, this may be written as
\begin{equation}
\label{eq: initial state}
|\Psi_n\rangle = \frac{1}{\sqrt{n!}} \left[ \frac{1}{\sqrt{c}}\int_{-\infty}^{\infty} dz~h(-z/c) \hat{\mathcal{E}}^{\dagger}(z, 0)  \right]^n \times \int dz~\mathcal{C}(z) \hat{C}^{\dagger}(z, 0) |0\rangle.
\end{equation}
The expectation value of $\hat{\rho}(x, y, t)$ with this state, denoted by $\rho_n (x,y, t) = \langle \Psi_n| \hat{\rho}(x,y, t) | \Psi_n \rangle$, then defines the density matrix of the stored spin wave  and forms our main quantity of interest.

Since $\partial_t\hat{\rho}(z, z)=0$, eqs.~(\ref{eq: EIT equation 1})-(\ref{eq: EIT equation 3}) can be solved straightforwardly in frequency space. Omitting irrelevant terms that depend on the vacuum initial operators $\hat{\mathcal{E}}(z,0)$, $\hat P(z,0)$, and $\hat S(z,0)$, the solution to the Rydberg spin wave operator can be written as (see  \ref{appa})
\begin{equation}\label{eq:SE}
\hat{S}(z, t) = \int dt^{\prime} \hat{e}(z, t - t^{\prime}) \hat{\mathcal{E}}(0, t^{\prime}),
\end{equation}
with the operator $\hat{e}(z, t - t^{\prime})$ to be discussed below.
Substituting this general solution for $\hat{S}(z, t)$ into Eq.~(\ref{eq: Density matrix equation of motion}) and taking expectation values with respect to $|\Psi_n\rangle$, we obtain the following equation of motion for the spin wave density matrix
\begin{align}
\label{eq: Density matrix equation of motion a}
i \partial_t \rho_n(x, y, t) =& ~ \frac{n}{c} \int dz \left[ V(z-y) - V(z-x) \right]  \int dt^{\prime} h^*(t^{\prime})  \int dt^{\prime \prime} h(t^{\prime \prime}) \nonumber\\
& \times   \langle \Psi_{n-1}| \hat{e}^{\dagger}(z, t\!-\!t^{\prime}) \hat{\rho}(x, y, t) \hat{e}(z, t\!-\!t^{\prime \prime}) |\Psi_{n-1}\rangle,
\end{align}
where we have used the property $\hat{\mathcal{E}}(0, t) |\Psi_n\rangle = \hat{\mathcal{E}}(-c t, 0) |\Psi_n\rangle = h(t) \sqrt{n/c}|\Psi_{n-1}\rangle$ of the target photons prior to entering the Rydberg medium.

In general, the operator $\hat{e}^{\dagger}(z, t)$ features a nonlinear dependence on the stored spin wave density operator $\hat{\rho}(z, z)$. However, when considering $\hat{e}(z, t) |\Psi_n\rangle$ in Eq.~(\ref{eq: Density matrix equation of motion a}), one can exploit the single occupancy of the stored mode $\hat{C}^{\dagger}(z)$ to simplify the problem. Upon normal ordering of the spin wave operators $\hat{C}(z)$ inside $\hat{e}(z,t)$, one is left with a linear $\hat{\rho}(x, y)$-dependence, since all higher order terms give vanishing contributions when applied to $|\Psi_n\rangle$. One can, hence, linearize $\hat{e}(z, t)$ according to $\hat{e}(z, t) = e_0(z, t) \mathbb{1} + \int dz^{\prime} e_1(z, z^{\prime}, t) \hat{\rho}(z^{\prime}, z^{\prime})$, where $e_0(z,t)$ and $e_1(z, z^{\prime},t)$ are complex valued coefficients whose explicit forms are derived in \ref{appa}. This procedure forms the key conceptual step in our derivation and can be straightforwardly extended to more complex $N$-body spin wave states, or coherences between different numbers of spin waves, by retaining higher order terms, as outlined in \ref{appb}. With the linearized expression for $\hat{e}(z, t)$ in the current context, the equation of motion for $\rho_n (x,y, t)$ may ultimately be written in the following manner
\begin{equation}
\label{eq: recursion relation}
\partial_t \rho_n (x,y, t) = n \Phi(x, y, t)\rho_{n-1} (x,y, t),
\end{equation}
where 
\begin{align}
\label{eq: Phi}
\Phi(x, y, t) = & ~ \frac{i}{c} \int dz \left[ V(z-x) - V(z-y)  \right]  \int dt^{\prime} h^*(t^{\prime})  \int dt^{\prime \prime} h(t^{\prime \prime}) \nonumber\\
& \times \left[ e_0^*(z, t-t^{\prime}) +  e_1^*(z, x, t-t^{\prime}) \right]  \left[ e_0(z, t-t^{\prime \prime}) +  e_1(z, y, t-t^{\prime \prime}) \right].
\end{align}
With the initial condition $\rho_0(x, y) = \mathcal{C}^*(x)\mathcal{C}(y)$, being the pure state of the initial density matrix, the full hierarchy of equations resulting from Eq.~(\ref{eq: recursion relation}) can be solved recursively in $n$ to finally yield
\begin{equation}
\label{eq: rho_n solution}
\begin{split}
\rho_n(x, y, t) & = \left[ 1 + \int_0^{t} d\tau \Phi(x, y, \tau) \right]^n \rho_0(x, y), \\
& = \left[ \frac{\rho_1(x, y, t)}{\rho_0(x,y)}\right]^n \rho_0(x,y).
\end{split}
\end{equation}
This result shows that all incident photons decohere the stored spin wave in an identical fashion, so the overall effect is the same whether the photons arrive simultaneously or sequentially. Physically, this linearity follows from the fact that photons only interact with the stored spin wave density, which is a static quantity, such that there is no effective interaction mediated between the target photons themselves. 

\begin{figure}[t!]
\hspace*{2.48cm}\includegraphics[width=0.84\columnwidth]{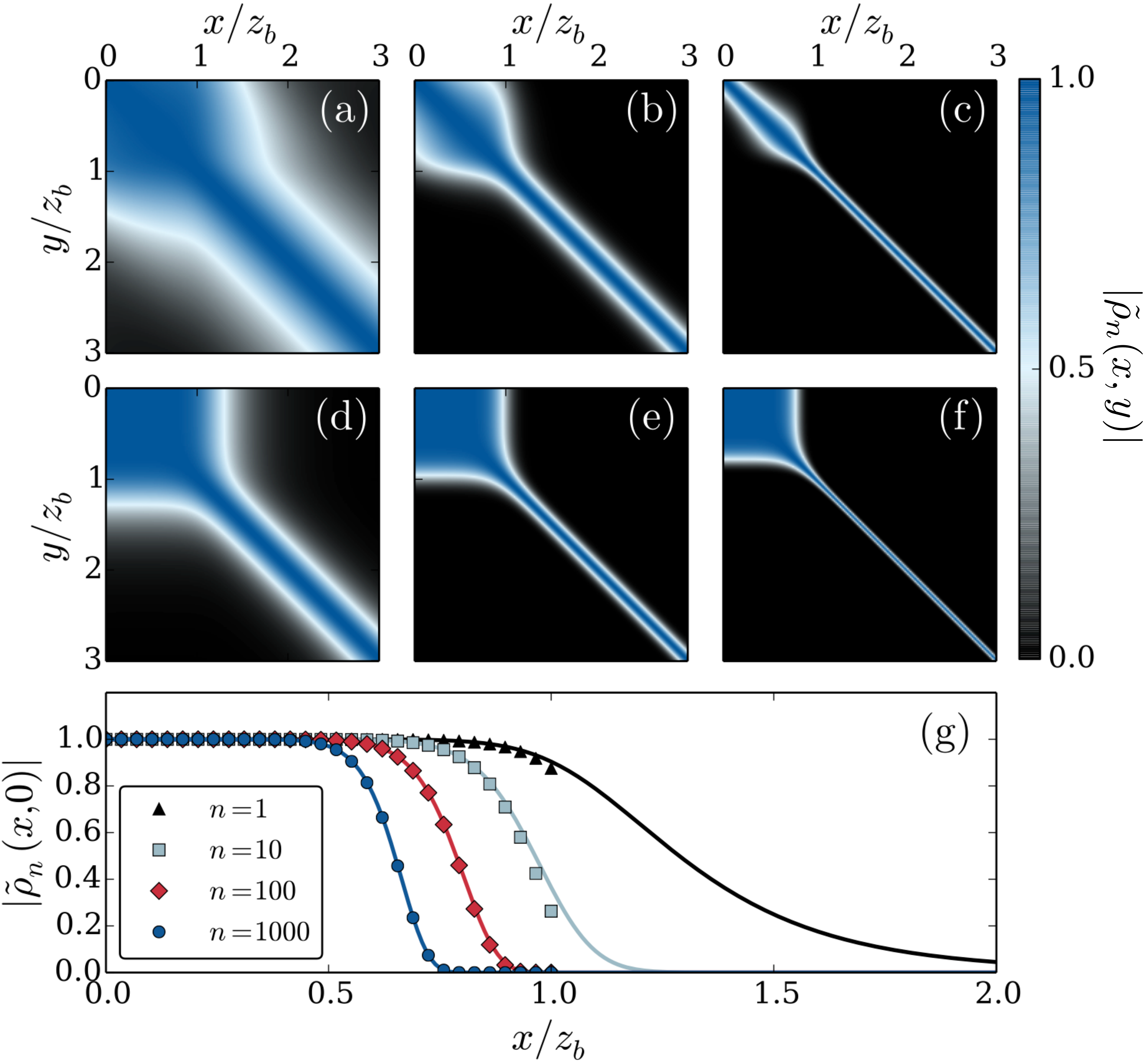}
\caption{\label{fig: Density matrices} (a), (b) and (c) show the rescaled final state of the stored spin wave density matrix $|\tilde{\rho}_n(x, y)| = |\rho_n(x,y,t\rightarrow \infty)/\rho_0(x,y)|$ after having interacted with $n=1, 10$ and $100$ target photons respectively for the case of $d_b=1$. (d), (e) and (f) show the corresponding behavior for $d_b=10$. (g) The profile of the coherent boundary feature along $|\tilde{\rho}_n(x, 0)|$ is shown for various indicated values of $n$ at $d_b=10$, comparing the numeric results (solid lines) to the approximate analytic solution for $x<z_b$ (points), according to Eq.(\ref{eq: analytic solution}).}
\end{figure}

To proceed, we numerically solve Eqs.~(\ref{eq: EIT equation 1})-(\ref{eq: Density matrix equation of motion}) to obtain the density matrix dynamics of the stored spin wave for the case of a single incoming target photon. Knowing $\rho_1$, Eq.~(\ref{eq: rho_n solution}) immediately yields the density matrix evolution for any $n$-photon Fock state. In Fig.~\ref{fig: Density matrices}(a-f) we show the final density matrix $\tilde{\rho}_n(x, y) = \rho_n(x, y, t\to\infty)/\rho_0(x, y)$ for different values of $n$ and $d_b$.

A universally observed feature in Fig.~\ref{fig: Density matrices}(a-f) is the pronounced loss of coherence beyond a blockade radius from the incident boundary, which turns the initial spin wave into a near classical distribution of the stored Rydberg excitation. This originates from projective position measurements of the stored excitation due to the spatially dependent nature of the photon scattering. For a scattering event occurring at a position $z>z_b$ in the medium, the stored excitation is projected to a region around $z+z_b$, as absorption most likely occurs one blockade radius away from the position of the stored excitation. Thus, the initially pure spin wave state is eventually decohered into a statistical mixture of localized excitations, as reflected by the narrow diagonal stripe in Fig.~\ref{fig: Density matrices}(a-f). 

The finite range, $z_b$, of the photon-spin wave interaction, however, offers a certain level of decoherence protection for the portion of spin wave within a blockade radius from the incident boundary. This is because an excitation stored in this region will cause photon scattering right at the medium boundary, irrespective of its exact location. Such immediate scattering therefore provides little spatial information about the stored spin wave state over this region, thereby causing less spatial decoherence.
However, in response to many repeated scattering events, this protection from decoherence is sensitively dependent on the optical depth of the medium. In particular, for $d_b\lesssim1$, one observes that the initial portion of the spin wave decoheres fairly quickly with an increasing number, $n$, of incident target photons [see Fig.~\ref{fig: Density matrices}(a-c)]. This is due to the fact that, in this limit, the absorption length is larger than the blockade radius, so there is an appreciable chance for a given photon to survive the dissipative interaction with the stored excitation. The extent of the amplitude attenuation suffered by a transmitted photon can then be significantly less than the expected amount of $\approx \exp [ - 2 d_b ]$ if the stored excitation is located near the medium boundary, since the length of the exposed effective two-level medium can be less than $2z_b$. This provides spatial information about the stored spin wave over $z \in [0, z_b]$, thus accounting for the eventual decoherence observed near the medium boundary with increasing $n$. On the other hand though, at large blockaded optical depth, $2d_b\gg1$, where the absorption length is much shorter than $z_b$, photons scatter over a much shorter length scale upon entry into the medium, so cannot probe the excitation position over a propagation depth $\sim z_b$. As such, the initial portion of the spin wave then remains more robust to decoherence with increasing $n$, as shown in Fig.~\ref{fig: Density matrices}(d-f). 

We can gain additional insights into this large-$d_b$ limit from an approximate analytical solution of the scattering dynamics for $x,y<z_b$ for long target pulses. In this limit, we can evaluate the static values of $e_0(z, t)$ and $e_1(z, z^{\prime}, t)$, for which one finds
\begin{equation}\label{eq:static}
e_0(z,t) + e_1(z, x,t) = i \frac{G}{\Omega} \frac{(z-x)^6}{z_{\rm b}^6 - i(z-x)^6}  \exp\!\left[ -\frac{G^2}{c \gamma} \int_0^z \frac{z_{\rm b}^6{\rm d}z^\prime}{z_{\rm b}^6 - i (z^{\prime} - x)^6} \right]\delta(t).
\end{equation}
Using this expression, eq.(\ref{eq: Phi}) can be solved approximately to yield 
\begin{equation}
\label{eq: analytic solution}
\tilde{\rho}_n(x, y)\approx [1 - (x^6 - y^6)^2/8z_b^{12}]^n,
\end{equation}
which agrees well with the numerical results, as shown Fig.~\ref{fig: Density matrices}(g). This indicates that
\begin{equation}
\label{eq: bscale}
N\approx 8(z_b/x)^{12} \gg 1
\end{equation}
scattered photons are required to decohere a spin wave component located at a distance $x<z_b$ from the entrance to the medium. Remarkably, this result is independent of $d_b$ and depends only on the shape of the potential, which implies that there is a fundamental limit in the protection to decoherence that is available by increasing $d_b$. This limit exists since the blockade is imperfect (i.e.\ the medium deviates from a two-level medium) any nonzero distance away from a stored $|r'\rangle$ excitation, so that the imaginary part of the susceptiblity at the entrance into the medium  -- and hence the absorption length of the incoming target photons -- depends on the position of the $|r'\rangle$ excitation. Similarly, we can also derive the width of the diagonal feature, which is found to scale with $d_b$ as $\sim 1/d_b^{5/11}$, indicating stronger decoherence beyond the boundary region with increasing $d_b$.

\section{Optimised Switching Protocol}
Having understood the many-body decoherence dynamics of the system, we are now in a position to optimize the entire switching protocol involving storage, decoherence and retrieval. Firstly, assuming that the incident gate photon is contained in a temporal mode $h_{\text{g}}(t)$, the initial storage can be analytically solved \cite{Gorshkov2007, Gorshkov2007a} to give
\begin{equation}
\label{eq: Optimal spin wave mode}
\mathcal{C}(z) = -\sqrt{\frac{d}{\gamma L}} \int_0^T dt \Omega_{\text{g}} e^{- zd/L - \Omega_{\text{g}}^2(T-t)/\gamma} I_0\left(2 \sqrt{zd \Omega_{\text{g}}^2(T-t)/L\gamma}\right)h_{\text{g}}(t),
\end{equation}
where $\mathcal{C}(z)$ is again the spatial profile of the stored spin wave. Here, $I_0(x)$ is the zeroth-order modified Bessel function of the first kind. Without loss of generality, we have assumed a square control field pulse of duration $T$ and constant Rabi frequency $\Omega_{\text{g}}$ which  facilitates the gate-photon storage. The density matrix of the stored spin wave after having interacted with an $n$-photon pulse can then be expressed according to Eq.~(\ref{eq: rho_n solution}) as $\tilde{\rho}_n(x, y) \mathcal{C}^*(x) \mathcal{C}(y)$. Finally, the efficiency $\eta$ of retrieving the stored gate photon in the backward direction, which is shown to be the optimal strategy \cite{Gorshkov2007, Gorshkov2007a}, can be written as
\begin{equation}
\label{eq: eta}
\eta = \int_0^L dz \int_0^L dz^{\prime} \frac{d}{2 L} \exp \left[- \frac{d}{2L} (z + z^{\prime}) \right] I_0\left(\frac{d}{L}\sqrt{zz^{\prime}}\right) \tilde{\rho}_n(z, z^{\prime}) \mathcal{C}^*(z) \mathcal{C}(z^{\prime}).
\end{equation}
Since Eq.~(\ref{eq: Optimal spin wave mode}) already includes the imperfect storage efficiency, Eq.~(\ref{eq: eta}) is in fact the total fidelity of the switch, taking into account photon storage, spin wave decoherence and retrieval, and can be readily optimized using power iteration methods \cite{Gorshkov2007, Gorshkov2007a}. Specifically, this procedure yields the optimal mode shape of the gate field $h_{\text{g}}(t)$ required to achieve storage into the optimal spin wave mode for a given $d_b$. We remark that, provided the duration $T$ of the control field is sufficiently long to store the entire length of the probe field $h_{\text{g}}(t)$, the optimisation is independent of $\Omega_{\text{g}}$ and the optimal storage solution can always be found by choosing $h_{\text{g}}(t)$ accordingly.
Note that the overall switching fidelity further depends on the probability (see \ref{appc})
\begin{equation}\label{eq:psc}
p_{\rm sc}=\exp\!\left(-2d_{\rm b}z_{\rm b}^{11}\int_{-z^\prime}^\infty{\rm d}x\int{\rm d}z^\prime\frac{\rho(z^\prime,z^\prime)}{z_{\rm b}^{12}+x^{12}}\right)
\end{equation}
to scatter a single photon off the stored gate excitation. Since $1-p_{\rm sc}$, thus, ranges between $\sim{\rm e}^{-4d_{\rm b}}$ and $\sim{\rm e}^{-2d_{\rm b}}$, the efficiency $\eta$, however, exponentially approaches the switch operation fidelity with increasing values of $d_{\rm b}$.

\begin{figure}[!t]
\hspace*{2.48cm}\includegraphics[width=0.84\columnwidth]{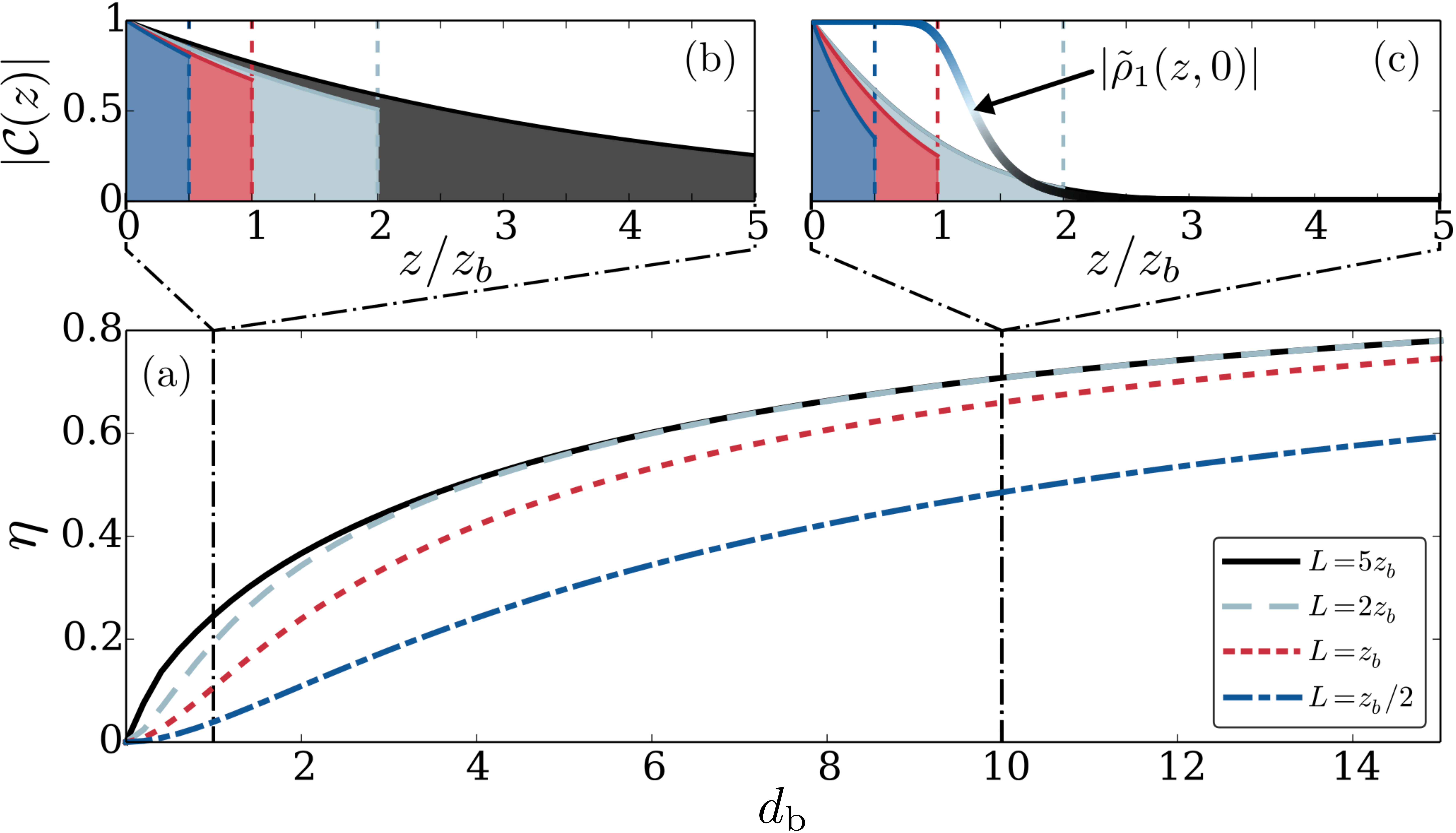}
\caption{\label{fig: eta ODb} (a) The combined efficiency $\eta$ of storage and retrieval is shown as function of $d_b$ for various indicated values of the medium length $L$. The mode profiles of the (unnormalized) optimally stored spin wave for $d_b=1$ and $10$ are shown in (b) and (c).}
\end{figure}

In Fig.~\ref{fig: eta ODb}(a), we show the efficiency, $\eta$, as a function of $d_b$ for the case of a single incident target photon, and for various medium lengths, $L$. Figs.~\ref{fig: eta ODb}(b) and (c) display the corresponding profiles of the optimal stored spin wave states at $d_b=1$ and $10$, respectively. From the above discussion one would naively expect that photon storage in a short medium of length $L\sim z_b$ is the universally optimal strategy \cite{Li2015a,Gorniaczyk2015}, since the photon is most protected from decoherence in this case. As evident from Fig.~\ref{fig: eta ODb}, this is, however, not the case, since at low $d_b\lesssim1$ the optimal stored spin wave profile $\mathcal{C}(z)$ turns out to be considerably longer than $z_b$. This is because the total optical depth for a short medium with $L\sim z_b$ and a small $d_b\lesssim1$ is not sufficient to provide for efficient storage and retrieval even in the absence of any spin wave decoherence. The optimal strategy is thus to find a compromise between minimising decoherence, by storing into a short medium, whilst maximising storage and retrieval efficiency by making the gate spin wave longer, despite then suffering from increased decoherence beyond a distance $z>z_b$ from the incident boundary [see Fig.~\ref{fig: eta ODb}(b)]. Only at larger $d_b$ [see Fig.~\ref{fig: eta ODb}(c)], where the blockaded boundary region provides for sufficient optical depth, does the straightforward strategy of storing into a short medium apply. Here, the optimal spin wave mode is observed to largely fit inside the profile, $|\tilde{\rho}_n(x, 0)|$, of the low-decoherence region of medium. As shown in Fig.~\ref{fig: eta ODb}(a), $\eta$ indeed no longer benefits from increasing the medium length beyond $L\approx 2z_b$ for large $d_b$. Related experiments currently realize values of $d_b\sim1$, which are largely limited by broadening effects \cite{Baur2014,Gaj2014} caused by additional spin wave dephasing at higher densities. With this current limitation on $d_b$, working with a small medium of length $L\sim z_b$ \cite{Li2015a} does therefore not present the optimal strategy for switching under experimentally relevant conditions.

\begin{figure}[!t]
\includegraphics[width=\columnwidth]{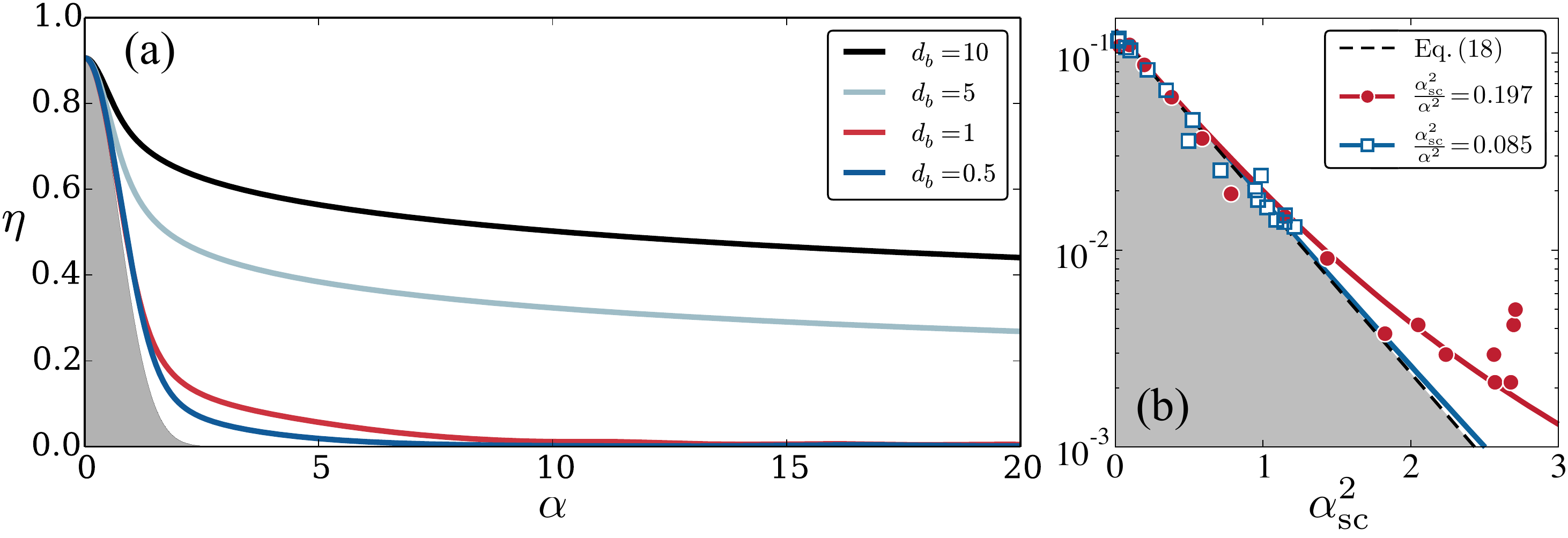}
\caption{\label{fig: eta alpha} (a) Storage and retrieval efficiency as a function of the incident target field amplitude $\alpha$ for various indicated values of $d_b$ and $d=50$. The shaded area indicates the contribution from the decreasing vacuum component of the target field, $|\langle n=0|\alpha\rangle|^2={\rm e}^{-\alpha^2}$. (b) Measured (symbols) \cite{Gorniaczyk2015} and calculated (lines) storage and retrieval efficiencies as a function of the number, $\alpha_{\rm sc}^2$, of scattered target photons for $\alpha_{\rm g}^2=0.5$ \cite{Hofferberth} and different, measured values of $\alpha_{\rm sc}^2/\alpha^2$. The solid lines show the results of our Monte Carlo simulations for the indicated parameters and the dashed line follows from Eq.~(\ref{eq:eta_exp}), which does not depend on $\alpha_{\rm sc}^2/\alpha^2$.}
\end{figure}

Present experiments typically do not operate with well defined photonic Fock states but use coherent input fields, i.e. multi-photon coherent states of light $|\alpha\rangle = \sum_n \frac{\alpha^{n}}{\sqrt{n!}}e^{-|\alpha|^2 / 2}|n\rangle$, containing an average number of $|\alpha|^2$ photons. The final density matrix of the spin wave state after decoherence due to its interaction with such a coherent target pulse can be straightforwardly obtained as 
\begin{equation}\label{eq:sw_coh}
\tilde{\rho}^{(\text{coh})}_{\alpha}(x,y) = \exp \left[ |\alpha|^2 \left( \tilde{\rho}_{n=1}(x,y)  -  1 \right)  \right]
\end{equation}
from the Fock-state results presented above\footnote{In the following we can, therefore, assume $\alpha$ to be real without loss of generality.}. In Fig.~\ref{fig: eta alpha}(a), we show the characteristic target-photon number dependence of the efficiency for different values of $d_b$. Common to all cases, one finds a rapid initial decrease of $\eta$. For small values of $d_b$, multi-photon scattering continues to diminish the spin wave coherence [see Fig.\ \ref{fig: Density matrices}(a-c)] such that the efficiency quickly vanishes as $\alpha$ is increased over the depicted interval. At larger values of $d_b$, however, decoherence protection in the boundary region becomes more robust against the scattering of multiple photons [see Fig.\ \ref{fig: Density matrices}(d-f) and Eq.\ (\ref{eq: bscale})] such that the efficiency decays only very slowly as the target-photon number is increased beyond $\alpha^2\sim1$. This large-$\alpha$ behaviour emergence from the weak dependence of the decoherence protection length on the target photon number, found in Eq.~(\ref{eq: bscale}).
In this regime, a strong increase of the target field intensity only marginally affects the retrieval efficiency, and thereby enables high-gain photon switching with little reduction of the overall operation fidelity.

The rapid initial drop of $\eta$ can be universally accredited to the decreasing vacuum component $|\langle n=0|\alpha\rangle|^2={\rm e}^{-\alpha^2}$ of the target pulse as indicated by the grey shaded region in Fig.~\ref{fig: eta alpha}(a). For small values of $\alpha$ we can thus employ a simplified picture assuming that all scattered target photons entirely inhibit gate retrieval, which in turn permits to straightforwardly extend the theory to arbitrary numbers of gate excitations. As described in \ref{appc}, the storage and retrieval efficiency can then be obtained from a simple Monte Carlo sampling of the scattering process. If we take the gate spin wave to be a coherent state with an average number of $\alpha_{\rm g}^2$ excitations, then for small values of the average number of scattered target photons, $\alpha_{\rm sc}$, the storage and retrieval efficiency is found to follow a simple exponential decay law
\begin{equation}\label{eq:eta_exp}
\eta=\eta_0{\rm e}^{-\alpha_{\rm sc}^2/\alpha_{\rm g}^2}
\end{equation}
where $\eta_0$ is the storage and retrieval efficiency without scattering. Recent experiments \cite{Gorniaczyk2015} have measured the storage and retrieval efficiency for different values of $z_{\rm b}$ (or equivalently different values of $\alpha_{\rm sc}^2/\alpha^2$) and reported a universal exponential decay as a function of $\alpha_{\rm sc}^2$. Eq.~(\ref{eq:eta_exp}) explains this universal behaviour and, for the measured value of $\alpha_{\rm g}^2=0.5$ \cite{Hofferberth}, quantitatively agrees with the experiment. As shown in Fig.~\ref{fig: eta alpha}(b), our corresponding Monte Carlo results reproduce the observed efficiencies even over the entire range of applied target field intensities. With the high level of quantitative agreement, our Monte Carlo approach also offers a new understanding of the observed deviations from Eq.~(\ref{eq:eta_exp}). In fact, the enhanced efficiency can be traced back to mutual decoherence protection by multiple gate excitations, whereby photon scattering off one excitation then prevents decoherence of subsequent excitations.
\section{Conclusion}
In summary, we have presented a many-body theory of spin wave decoherence in a single photon switch based on Rydberg-EIT. This has been used to work out an optimal switching protocol and to determine maximum achievable switching fidelities for a given set of all relevant experimental parameters. The presented results are, thus, of direct relevance to ongoing transistor experiments \cite{Baur2014, Gorniaczyk2014, Tiarks2014, Gorniaczyk2015}, while the developed theoretical framework can be applied to a range of other quantum optical applications involving photon storage in Rydberg media \cite{Pohl2010, Honer2011, Gorshkov2011, Otterbach2013, Tiarks2016} and permits straightforward extensions to more complex many-body states of gate and target photons. 

The optimal cloud dimensions where shown to be sensitive to the available Rydberg atom interactions and atomic densities, i.e.\ the achievable optical depth, $2d_b$, per Rydberg blockade radius. While short optical media provide for the highest coherence protection, it turns out that choosing a short medium just covering a single blockade radius \cite{Li2015a} is not universally optimal and particularly not under conditions of current experiments for which $d_b\sim1$ \cite{Baur2014,Gorniaczyk2014,Tiarks2014,Gorniaczyk2015}. This unexpected behaviour was shown to arise from both the effects of multiple photon scattering as well as the interplay between interaction-induced decoherence and the gate field dynamics during storage and retrieval, not considered in previous work \cite{Li2015a}.

By extending the presented theory to multiple gate excitations, we have provided a new understanding and accurate description of recent measurements \cite{Gorniaczyk2015} of storage and retrieval efficiencies for photon switching in a cold rubidium Rydberg gas. Our results show that the observed efficiency is largely dominated by the vacuum component of the incident target field, but also reveal a new decoherence protection mechanism that emerges for multiple gate excitations. We remark that the difference to the interpretation suggested in \cite{Gorniaczyk2015} is rooted in the coherent-state nature of the gate photons and their finite storage fidelity, disregarded in the theoretical analysis of \cite{Gorniaczyk2015}.

We finally note that the dissipative nature of the switching mechanism in the current context fundamentally restricts applications to the domain of classical switching. Anticipated quantum applications \cite{Baur2014,Gorniaczyk2014,Li2015a} are inherently precluded by target photon scattering, since this fully decoheres any quantum superposition involving the vacuum component of the stored gate excitation, even when its spatial coherence can be completely preserved. Extensions into the quantum regime require to control the mode into which target photons are scattered, amounting to a coherent switching mechanism. Aside from enabling true quantum applications, this would also eradicate scattering induced spin wave decoherence, allowing storage and retrieval to benefit from the total optical depth of the entire medium, and, thereby, making efficient switching possible at much lower values of $d_b$. Achieving such a coherent nonlinearity will likely require hybrid  architectures offering strong mode confinement \cite{Chen2013,Das2016} or new schemes altogether.

\ack
We thank H.~P.~B\"uchler, O.~Firstenberg, M.~Gullans, S.~Hofferberth, I.~Lesanovsky, M.~Maghrebi, I.~Mirgorodskiy, Y.~Wang and E.~Zeuthen for helpful discussions. We are particularly grateful to Sebastian Hofferberth and Hannes Gorniaczyk for sharing data of the experiment \cite{Gorniaczyk2015} and valuable discussions on their measurements. This research was supported by  ARL CDQI, NSF PFC at JQI, NSF QIS, AFOSR, ARO, ARO MURI, as well as by the EU through the FET-Open Xtrack Project HAIRS and the FET- PROACT Project RySQ.

\appendix
\section{Derivation of $\hat{e}(z,z^\prime,t)$}\label{appa}
Below we outline the solution to the dynamics of the spin wave operator $\hat{S}(z, t)$ for the case in which a single gate excitation has been stored in the medium. We start by Fourier transforming the Heisenberg Eqs.~(\ref{eq: EIT equation 1}-\ref{eq: EIT equation 3}) to obtain a set of equations for $\tilde{\mathcal{E}}(z, \omega)  = (\sqrt{2 \pi})^{-1/2} \int_{\infty}^{\infty} d t e^{i\omega t} \hat{\mathcal{E}} (z,t)$, etc.. Again, this is straightforward since the stored spin wave density operator $\hat{\rho}(z^{\prime}, z^{\prime})$ is time independent. Solving for $\tilde{P}(z, \omega)$ one obtains a closed set of equations 
\begin{align}
\label{eq: Fourier equation of motion for E}
c \partial_z \tilde{\mathcal{E}}(z, \omega) & = i\omega \tilde{\mathcal{E}}(z, \omega) - i \frac{G^2}{\omega + i \gamma} \tilde{\mathcal{E}}(z, \omega) -i \frac{G\Omega}{\omega + i \gamma} \tilde{S}(z, \omega), \\
\label{eq: Fourier equation of motion for S}
\omega\tilde{S}(z, \omega) & = \frac{G\Omega}{\omega + i \gamma} \tilde{\mathcal{E}}(z, \omega) + \frac{\Omega^2}{\omega + i \gamma} \tilde{S}(z, \omega) + \int dz^{\prime} V(z-z^{\prime}) \hat{\rho}(z^{\prime}, z^{\prime})  \tilde{S}(z, \omega), 
\end{align}
for the photon and target spin wave operators. The latter can then be expressed as
\begin{equation}
\label{eq: S solution}
\tilde{S}(z, \omega) = -\frac{G\Omega}{\Omega^2 - \omega(\omega + i \gamma)} \frac{\tilde{\mathcal{E}}(z, \omega)}{1 + \int dz^{\prime} U(z-z^{\prime}, \omega) \hat{\rho}(z^{\prime}, z^{\prime})},
\end{equation}
where we have introduced the effective potential
\begin{equation}
U(z, \omega) = \left[\frac{\omega + i \gamma}{\Omega^2 - \omega (\omega + i \gamma)} \right] V(z).
\end{equation}
As mentioned in the main text, the general solution for $\tilde{S}(z, \omega)$ in eq.(\ref{eq: S solution}) is inherently nonlinear in the stored spin wave density $\hat{\rho}(z^{\prime}, z^{\prime})$. However, by expanding 
\begin{equation}
\frac{1}{1 + \int dz^{\prime} U(z-z^{\prime}, \omega) \hat{\rho}(z^{\prime}, z^{\prime})} = \sum_{k=0} (-1)^k \left[ \int dz^{\prime} U(z-z^{\prime}, \omega) \hat{\rho}(z^{\prime}, z^{\prime}) \right]^k,
\end{equation}
we can now make use of the fact that $|\Psi_n\rangle$ only contains a single stored excitation and retain only linear terms in $\hat{C}^\dagger(z)\hat{C}(z)$ after normal ordering the operators. It then follows that
\begin{equation}
\frac{1}{1 + \int dz^{\prime} U(z-z^{\prime}, \omega) \hat{\rho}(z^{\prime}, z^{\prime})} |\Psi_n\rangle= \left[1 - \int dz^{\prime}  
\frac{U(z-z^{\prime}, \omega)}{1 + U(z-z^{\prime}, \omega)}\hat{\rho}(z^{\prime}, z^{\prime})\right]|\Psi_n\rangle,
\end{equation}
and that the spin wave wave operator $\tilde{S}(z, \omega)$ can be written as
\begin{equation}
\label{eq: S solution 2}
\tilde{S}(z, \omega) = -\frac{G\Omega}{\Omega^2 - \omega(\omega + i \gamma)} \left[ 1 - \int dz^{\prime}  \frac{U(z-z^{\prime}, \omega)}{1 + U(z-z^{\prime}, \omega)}   \hat{\rho}(z^{\prime}, z^{\prime}) \right] \tilde{\mathcal{E}}(z, \omega). 
\end{equation}
Substitution into Eq.~(\ref{eq: Fourier equation of motion for E}) then yields a closed propagation equation for $\tilde{\mathcal{E}}(z, \omega)$ whose solution is
\begin{equation}
\label{eq: E solution}
\tilde{\mathcal{E}}(z, \omega) = \tilde{\mathcal{E}}(0, \omega) \exp\!\left[ i \chi_0(\omega) z - i \chi_V(\omega)  \int_0^z dx \int dz^{\prime}  \frac{U(x-z^{\prime}, \omega)}{1 + U(x-z^{\prime}, \omega)} \hat{\rho}(z^{\prime}, z^{\prime})   \right],
\end{equation}
where we have introduced the quantities
\begin{align}
\chi_0(\omega) & = \frac{1}{c} \left[ \omega + \frac{G^2 \omega}{\Omega^2 - \omega(\omega + i\gamma)}  \right], \\ 
\chi_V(\omega) & = \frac{1}{c} \frac{G^2 \Omega^2}{\left[\omega + i\gamma\right] \left[\Omega^2 - \omega(\omega + i\gamma)\right]}.
\end{align}
Here, $\chi_0(\omega)$ is the optical susceptibility of the EIT medium in the absence of interactions, whilst  $\chi_0(\omega) - \chi_V(\omega)$ is that of a resonant two-level medium with the Rydberg state blocked. Expectedly, for large distances $|z - z^{\prime}|$ between a target photon and the stored gate excitation, the photons thus experience an EIT medium, whilst for small distances  $|z - z^{\prime}|<z_{\rm b}$ they experience an effective two-level medium with enhanced absorption. In the limit of long target pulses, Eq.~(\ref{eq:psc}) simply follows from Eq.~(\ref{eq: E solution}) as $p_{\rm sc}=1-|\tilde{\mathcal{E}}(\infty, 0)|^2/|\tilde{\mathcal{E}}(0, 0)|^2$

Linearising again the nonlinear $\hat{\rho}(z^\prime,z^\prime)$-dependence in Eq.~(\ref{eq: E solution}), substituting the result into Eq.~(\ref{eq: S solution}), and Fourier transforming back to the time domain then yields the desired expression Eq.~(\ref{eq:SE}), where the Fourier transforms of the complex coefficients $e_0(z,t)$ and $e_1(z,z^{\prime},t)$ are explicitly given by
\begin{align}
\label{eq:e0}
\tilde{e}_0(z, \omega) & = -\frac{G\Omega}{\Omega^2 - \omega(\omega + i \gamma)} \exp\left[ i \chi_0(\omega) z \right],  \\
\label{eq:e1}
\tilde{e}_1(z, z^{\prime}, \omega) &  =   \left(\frac{ \exp\!  \left[ -i\chi_V(\omega) \int_0^z dx  \frac{U(x-z^{\prime}, \omega)}{1 + U(x-z^{\prime}, \omega)}  \right]}{1 + U(z-z^{\prime}, \omega)}  - 1 \right)  \tilde{e}_0(z, \omega).
\end{align}
Eq.~(\ref{eq:static}) then simply follows from Eqs.(\ref{eq:e0}) and (\ref{eq:e1}) by setting $\omega=0$. Let us finally use this result to derive the scaling relation Eq.~(\ref{eq: analytic solution}). Substitution into Eq.~(\ref{eq: Phi}) and carrying out the time integration yields
\begin{align}
\int_0^\infty{\rm d}t \Phi(x, y, t) = & ~ i d_{\rm b} \int dz z_{\rm b}^5 \frac{(z-y)^6-(z-x)^6}{[z_{\rm b}^6 + i(z-x)^6][z_{\rm b}^6 - i(z-y)^6]}  \nonumber\\
& \times \exp\!\left[ -d_{\rm b} \int_0^z {\rm d}z^\prime \left(\frac{z_{\rm b}^5}{z_{\rm b}^6 + i (z^{\prime} - x)^6}+ \frac{z_{\rm b}^5}{z_{\rm b}^6 - i (z^{\prime} - y)^6}\right) \right].
\end{align}
For $d_{\rm b}\gg1$ and $x,y<z_{\rm b}$ the exponential function is sharply peaked around $z=0$, such that we can evaluate the exponent for $z^\prime=0$ and set $z=0$ everywhere else. The result 
\begin{align}
\int_0^\infty{\rm d}t \Phi(x, y, t) = & ~ i d_{\rm b} \int dz z_{\rm b}^5 \frac{y^6-x^6}{[z_{\rm b}^6 + ix^6][z_{\rm b}^6 - iy^6]}   \exp\!\left[ -d_{\rm b} \left(\frac{z_{\rm b}^5}{z_{\rm b}^6 + i x^6}+ \frac{z_{\rm b}^5}{z_{\rm b}^6 - i y^6}\right) z\right]\nonumber\\
= & ~  \frac{2z_{\rm b}^6 }{2z_{\rm b}^6 + i (x^6 -  y^6)}-1
\end{align}
is indeed independent of $d_{\rm b}$ and immediately leads to Eq.~(\ref{eq: analytic solution}).

\section{Two Stored Excitations}\label{appb}
To illustrate that our derivation can be straightforwardly extended to the case of $N$ stored excitations and to coherent superpositions between different $N$, we briefly remark in this Appendix on the case of $N= 2$. In this case, Eq.(\ref{eq: Density matrix equation of motion a}) becomes 
\begin{equation}
\begin{split}
& i \partial_t \langle \Psi_n | \hat{C}^{\dagger}(x_1,t) \hat{C}^{\dagger}(x_2,t) \hat{C}(y_1,t) \hat{C}(y_2,t) |\Psi_n\rangle =  \\
& = \frac{n}{c} \int d z [V(z-y_1) + V(z-y_2) - V(z-x_1) - V(z-x_2)]  \int dt^{\prime} h^*(t^{\prime})  \int dt^{\prime \prime} h(t^{\prime \prime})  \\
& ~~~ \times   \langle \Psi_{n-1}| \hat{e}^{\dagger}(z, t\!-\!t^{\prime}) \hat{C}^{\dagger}(x_1,t) \hat{C}^{\dagger}(x_2,t) \hat{C}(y_1,t) \hat{C}(y_2,t) \hat{e}(z, t-t^{\prime \prime}) |\Psi_{n-1}\rangle.
\end{split}
\end{equation}
Keeping now, inside $\hat{e}$, terms up to second order in $\hat{\rho}(z^{\prime},z^{\prime})$, one can reduce the right-hand-side to a recursive dependence on $\langle \Psi_{n-1} | \hat{C}^{\dagger}(x_1,t) \hat{C}^{\dagger}(x_2,t) \hat{C}(y_1,t) \hat{C}(y_2,t) |\Psi_{n-1}\rangle$, exactly as in Eq.\ (\ref{eq: recursion relation}). The resulting hierarchy of equations can be solved recursively to give a solution similar to Eq.\ (\ref{eq: rho_n solution}), confirming again that the incident photons decohere the stored state in an independent fashion.  

Any coherence between different numbers $N$ of stored excitations, e.g. between $N=1$ and $N=2$, $\langle \Psi_n| \hat{C}^{\dagger}(x_2,t) \hat{C}(y_1,t) \hat{C}(y_2,t) |\Psi_n\rangle$, or between $N=1$ and $N= 0$, $\langle \Psi_n | \hat{C}(y_1,t) |\Psi_n\rangle$, can be computed similarly.

\section{Retrieval Efficiency for coherent-state gate excitations} \label{appc}
The results of Fig.\ \ref{fig: eta alpha}(a) show that the initial drop of the retrieval efficiency largely reflects the decreasing vacuum component of the target photon pulse. This behaviour suggests a simplified picture based on the assumption that any scattered target photon completely inhibits retrieval, while the gate spin wave remains virtually unaffected by transmitted photons. 
Below we provide a derivation of Eq.~(\ref{eq:eta_exp}) using this idea.

We consider a coherent state, ${\rm e}^{-\alpha_{\rm g}/2}\sum_{n_{\rm g}}\frac{\alpha_{\rm g}^{n_{\rm g}}}{n_{\rm g}!}|n_{\rm g}\rangle$, of the stored gate spin wave with an average number of $\alpha_{\rm g}^2$ excitations. Then the storage and retrieval efficiency  
\begin{equation}\label{eq:eff}
\eta=\frac{\eta_0}{\alpha_{\rm g}^2}{\rm e}^{-\alpha^2}{\rm e}^{-\alpha_{\rm g}^2}\sum_{n,n_{\rm g}}\frac{\alpha^{2n}}{n!}\frac{\alpha_{\rm g}^{2n_{\rm g}}}{n_{\rm g}!}\sum_{n_{\rm g}^\prime\le n_{\rm g}}P^{(n)}_{{n_{\rm g}},{n_{\rm g}^\prime}} n_{\rm g}^\prime
\end{equation}
can be calculated from a coherent-state average of the surviving excitations, $n_{\rm g}^\prime$, over the number distribution of the target photons and gate excitations. Here, $\eta_0$ denotes the storage and retrieval efficiency in the absence of interactions ($\alpha=0$) and $P^{(n)}_{{n_{\rm g}},{n_{\rm g}^\prime}}$ is the probability that $n$ incident target photons scatter off $n_{\rm g}-n_{\rm g}^\prime$ out of $n_{\rm g}$ gate excitations. 
 
Under typical conditions of low excitation densities \cite{Baur2014, Gorniaczyk2014,Tiarks2014,Gorniaczyk2015}, we can assume that  the blockade volumes of different gate excitations do not overlap and calculate these probabilities in a sequential fashion with a single photon scattering probability $p_{\rm sc}$ per gate excitation. The probability to preserve all excitations then simply follows as
\begin{equation}
P^{(n)}_{{n_{\rm g}},{n_{\rm g}}}=(1-p_{\rm sc})^{n_{\rm g}n},
\end{equation}
while the probability to decohere exactly one excitation 
\begin{equation}
P^{(n)}_{{n_{\rm g}},{n_{\rm g}}-1}=\sum_{k=1}^{n_{\rm g}}\left[(1-p_{\rm sc})^{k-1}p_{\rm sc}+(1-p_{\rm sc})^{n_{\rm g}}\right]^{n}-(1-p_{\rm sc})^{n_{\rm g}n}
\end{equation}
is given by the cumulative probability to scatter of the $k$th excitation. To linear order in $p_{\rm sc}$ higher order terms do not contribute and we can write $P^{(n)}_{{n_{\rm g}},{n_{\rm g}}}\approx1-n_{\rm g}np_{\rm sc}$ and $P^{(n)}_{{n_{\rm g}},{n_{\rm g}}-1}\approx n_{\rm g} n  p_{\rm sc}$. Substituting these expressions into Eq.\ (\ref{eq:eff}) yields
\begin{align}
\eta&=\frac{\eta_0}{\alpha_{\rm g}^2}{\rm e}^{-\alpha^2}{\rm e}^{-\alpha_{\rm g}^2}\sum_{n,n_{\rm g}}\frac{\alpha^{2n}}{n!}\frac{\alpha_{\rm g}^{2n_{\rm g}}}{n_{\rm g}!}n_{\rm g}\left(1- n_{\rm s} p_{\rm sc}\right)\nonumber\\
&=\eta_0{\rm e}^{-\alpha^2}\sum_{n}\frac{\alpha^{2n}}{n!}\left(1- n_{\rm s} p_{\rm sc}\right)\approx\eta_0{\rm e}^{-\alpha^2}\sum_{n}\frac{\alpha^{2n}}{n!}\left(1- p_{\rm sc}\right)^n=\eta_0{\rm e}^{-p_{\rm sc}\alpha^2}.
\end{align}
By re-expressing $\alpha^2$ in terms of the average number $\alpha_{\rm sc}^2=\alpha^2(1-{\rm e}^{-p_{\rm a}\alpha_{\rm g}^2})\approx\alpha^2\alpha_{\rm g}^2p_{\rm a}$ of scattered photons we finally obtain Eq.~(\ref{eq:eta_exp}). 

In order to verify the involved small-$p_{\rm sc}$ expansion we also performed numerical calculations by on a random sampling of the scattering probabilities $P^{(n)}_{{n_{\rm g}},{n_{\rm g}^\prime}}$ and a Monte Carlo integration of the sums in Eq.~(\ref{eq:eff}). Only requiring $\eta_0$, $\alpha_{\rm sc}^2/\alpha^2$ and $\alpha_{\rm g}^2$ as input parameters, which are all known in the experiment of \cite{Gorniaczyk2015,Hofferberth}, the simulations reproduce the measured retrieval efficiencies remarkably well. Moreover, our Monte Carlo results perfectly match the small-$\alpha$ prediction Eq.~(\ref{eq:eta_exp}) for any considered combination of parameters. 

\section*{References}
\bibliography{references.bib}

\end{document}